\documentstyle[aps,twocolumn,epsf,rotate,prl]{revtex}
\newcommand{\be}{\begin{equation}}
\newcommand{\ee}{\end{equation}}
\newcommand{\bdm}{\begin{displaymath}}
\newcommand{\edm}{\end{displaymath}}
\newcommand{\ba}{\begin{eqnarray}}
\newcommand{\ea}{\end{eqnarray}}

\begin{document}

%\draft
%\twocolumn[\hsize\textwidth\columnwidth\hsize\csname @twocolumnfalse\endcsname

\title{
$\pi$ states in Josephson Junctions between $^3$He-B
}
\author{S.-K. Yip}
\address{
Physics Division, National Center for Theoretical Sciences ,
P. O. Box 2-131,  Hsinchu, Taiwan 300,
R. O. C.}
\date{\today}
\maketitle

\begin{abstract}
{\small

We investigate the dependence of the current-phase relationship
on the orientation of the order parameter for a pinhole
between two $^3$He-B reservoirs. We show that, due to
the internal spin structure of the superfluid, the
energy of the junction may have a relative minimum 
at phase difference equals $\pi$ at low temperatures.  The dependence
of the supercurrent on the direction of an applied
magnetic field can be used to verify the present
mechanism for the ``$\pi$-states''.

}

\end{abstract}
\date{\today}
\vspace*{0.2 cm}

Superfluid $^3$He is a remarkable state of matter.  It exhibits
two superfluid phases A and B in zero magnetic field.  In particular,
in the B-phase, the magitude of the gap is independent of
momentum direction $\hat p$ despite the fact that the pairing is triplet.
It does so by having $S = 1$ pairs with zero spin projections
 [ $ { 1 \over \sqrt{2} } (|\uparrow \downarrow > + \ 
 |\downarrow \uparrow >) $] along direction $\hat d (\hat p)$, 
with $\hat d$ related to $\hat p$ by a rotation:
$\hat d_i (\hat p) = R_{i \mu} (\hat n, \theta) \hat p_{\mu} $.
Here $R(\hat n, \theta)$ is the rotational matrix with
rotational axis $\hat n$ and angle $\theta$.
In the bulk $\theta$ is determined by minimizing the dipole
interaction energy and is given by 
$\theta_L \equiv {\rm cos}^{-1} (-{1 \over 4}) 
\approx 0.58 \pi$,
usually referred to as the Leggett angle. 
For any quantization axis perpendicular to $\hat d$,
there are only $ |\uparrow \uparrow> $ and $ |\downarrow \downarrow> $
pairs.  In the bulk the amplitudes for these pairs are 
equal, though the phase angles can be different. \cite{Leggett75}

  Being a superfluid, one naturally expects a Josephson current
can flow across weak-links between two reservoirs.
This was studied experimentally first by Avenel and Varquaux
\cite{Avenel88}.  
%Their junction consist of a slit on a thin
%membrane seperating the two $^3$He reservoirs.
Recently the Berkeley group have studied
this Josephson effect again in much more detail,
 in particular the current
phase relationships have been mapped out \cite{Backhaus98}.   
In their experiment the weak-links consist of a large
number of small apertures (diameter $\approx 0.1 \mu m$)
made on a thin membrane. A particular
interesting feature is the $\pi$-states which occur when
the temperature $T$ is not too close to $T_c$.
%where the energy of the junction possesses a relative
%minimum at phase difference equals to $\pi$.
For an ordinary junction between two
s-wave superconductors, the current-phase relationship
is a slanted sine function, with the current $I$ positive
for phase difference $\chi$ satisfying $0< \chi < \pi$. 
The junction energy $E$, related to the current $I$ by
$ I = dE/d\chi$, is maximum at $\chi = \pi$.
However, the current-phase relationships in \cite{Backhaus98} 
are slanted sines only at $T$ not too far from $T_c$.  
At lower temperatures,   though $I$ is
positive for small $\chi$ as usual, it turns negative at a phase
difference less than $\pi$, and $I$ passes through $0$ again
at $\chi = \pi$ with  $dI/d\chi > 0$. Thus $\pi$ is a relative
minimum instead of maximum in the junction  free energy.
Explanation of this phenomenon is still controversial.
There is a suggestion \cite{AMV} that this $\pi$ state is not
an intrinsic property of a single junction but
rather the  collective behavior of many pinholes.
Earlier theoretical works \cite{Thuneberg88,Hook87} have
also predicted possible existence of $\pi$-states,
but they rely on finite length or the width of
the channels.

Later experiments \cite{Packard99} reveal that 
there are in fact two possible current-phase relationships.  
They can be achieved from 
different cooldowns from the normal state.
 These two states are
distinguishable also by the different magnitudes of the
critical current.
% though both exhibit $\pi$-states at sufficiently low temperatures. 
 The authors of Ref \cite{Backhaus98,Packard99} suggested
that the two different states may be the result of two
different relative orientations of $\hat n$ on the two
sides of the junction; such as parallel or anti-parallel.
%(both along the normal to the surface).

The geometry of the individual apertures in
 the experiment of \cite{Backhaus98,Packard99}
 approaches that of pinholes, {\it i.e.},
apertures that have dimensions much less than
the coherence length ($\approx 0.1 \mu m$).
Though this criterion is not strictly obeyed for
the experiment,
pinholes are much easier to study theoretically,
since in this case all the self-consistent fields,
including superfluid pairing and Fermi liquid
effects, are the same as those near an impenetrable wall. \cite{KO77}
The current-phase relationship of a single pinhole
in $^3$He-B has been investigated by Kurkij\"arvi.\cite{Kurki88}  
He implicitly assumed that $\hat n$ on both sides of
the junction are parallel.  Ignoring surface depairing
he found that $I (\chi)$ is simply that of an s-wave
superfluid/superconductor.  We shall reconsider the
single pinhole junction, but allowing general relative
orientations of $\hat n^{l,r}$ on the left and right
of the junction.  In particular, we shall show that
one has a natural mechanism for the $\pi$-state if
$\hat n^{l}$ and $\hat n^{r}$ are not parallel.
The basic origin of the $\pi$-state is due to
the internal spin structure of the order parameter.
For a given momentum direction $\hat p$, quasiparticles
of different spin projections actually see different effective
phase differences across the junction; thus
contributing to $I (\chi)$ with a phase shifted from
each other. Provided $T$ is not too close to $T_c$, 
the resultant current-phase
relationship is anomalous, and in general 
can have an energy-phase relation that has  a relative
minimum at $\chi = \pi$.  Thus we have one of the  very unusual
situations where the spin structure of the order parameter
affects the dynamics of the mass flow.

The Berkeley experiments \cite{Backhaus98,Packard99} were 
performed in the absence of magnetic fields.  In this case
$\hat n$ prefers to lie along (or opposite to) the normal near
a surface thus of the membrane seperating the two 
reservoirs.  However, under
a magnetic field $\vec H$ its orientation can be modified. 
\cite{BC78} We shall
show that $I(\chi)$ can be changed substantially by
applying a magnetic field of sufficient magnitude
($ {> \atop \sim} 50 {\rm G}$) along a general direction.
  Moreover, we predict
that if one performs cool-downs from the normal state under a magnetic field
in general directions, one can have more than two 
current-phase relationships possible. These predictions
can be used to distinguish among the 
different hypotheses suggested for the $\pi$-state.

We shall then consider a single pinhole between
two reservoirs $l$ and $r$ with $\hat z$ being the
direction along the interface normal.  The current can
be calculated along the same lines as in Ref. 
\cite{KO77,Kurki88}.  
 Without loss in generality we
take the phase of the order parameter to be $0$ and $\chi$
for the left and right reservoirs respectively.
To obtain the current we need to solve the Andreev \cite{Andreev64}
equation (suitably generalized to triplet pairing) or
the quasiclassical Green's function.\cite{SR83}
For simplicity we shall ignore  
surface depairing.   Under this approximation
the problem simplifies enormously by the observation
that one only has $\vec S \cdot \hat w = \pm 1$ pairs
along any direction $\hat w$ perpendicular to $\hat d$.
$\hat n^{l,r}$ can be considered as constants in the present
calculations since the size of the pinhole is much 
less than that of the coherence length which is
in turn much less the bending length of the $\hat n$ vectors.
Thus for given $\hat p$ and thus quasiparticle path
through the pinhole, $\hat d$ is piecewise constant and
equals either $\hat d^{l}$ or $\hat d^{r}$. 
 By choosing
the spin-quantization axis along $\hat w \equiv
\hat d^l(\hat p) \times \hat d^r (\hat p)$,
the gap matrix is finite only for the $\uparrow \uparrow$
and $\downarrow \downarrow$ components.
Explicitly,  
  with the triad $(\hat u, \hat v, \hat w)_{\hat p}$
as the basis vectors for $\hat d (\hat p)$, the gap matrix
has the form 
\begin{displaymath}
{\bf \Delta} = \Delta_B 
\left(
\begin{array}{cc}
- d_u + i d_v & 0 \\
0 & d_u + i d_v 
\end{array}
\right)
\ = \ 
\Delta_B
\left(
\begin{array}{cc}
-e^{- i \phi_p} & 0 \\
0 & e^{i \phi_p}
\end{array}
\right)
\end{displaymath}
\noindent
where $\phi_p$ is the azimuthal angle of $\hat d$ 
in the $(u,v)$ plane. $\phi_p = \phi_p^l (\phi_p^{r})$
for $ z < (>) 0$.
The Andreev
equation or the quasiclassical equation block-diagonalized
in spin-space, resulting in two matrix equations only
in particle-hole space.  Each of them can be solved
as in the s-wave case. For given $\hat p$, an $\uparrow$ quasiparticle
 effectively sees a phase $ \pi- \phi_p^l$ for $z < 0$ and
$ \pi - \phi_p^r + \chi$ for $ z > 0$.
{\it i.e.}, an effective phase difference of 
$\chi - ( \phi_p^r - \phi_p^l)$.  Similarly
the effective phase difference for a $\downarrow$
quasiparticle is  
$\chi + ( \phi_p^r - \phi_p^l)$.
For future convenience we shall define $ \chi_{\hat p}^s
\equiv 
  \phi_p^r - \phi_p^l$.  Obviously $\chi_{\hat p}^s$
corresponds to the angle between $\hat d^r (\hat p)$
and $\hat d^l (\hat p)$, thus 
$\chi_{\hat p}^s = {\rm cos}^{-1} ( \hat d^r(\hat p) \cdot
\hat d^l(\hat p) ) $.  We see that the contribution
of the present quasiparticle path to the current
is proportional to the sum
\begin{equation}
\sum_{\sigma = \pm 1} 
\Delta_B {\rm sin} \left( {\chi - \sigma \chi_{\hat p}^s \over 2} \right ) 
{\rm tanh} \left(
{\Delta_B \over 2 T}
 {\rm cos} \left( {\chi - \sigma \chi_{\hat p}^s \over 2} \right ) 
\right)
\label{sum}
\end{equation}

\begin{figure}[h]
\centerline
{ \epsfxsize=0.22\textwidth
\epsfysize = 0.37\textwidth
\rotate[r]
{ \epsfbox{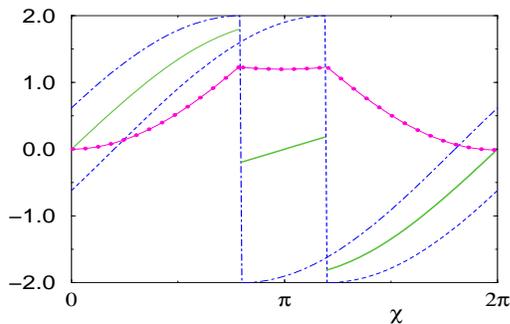} } }
\vskip 0.4 cm
\begin{minipage}{0.45\textwidth}
\caption[]{ The basic mechanism of producing a $\pi$-state.
Dashed and dot-dashed lines are two current-phase
relationships shifted from each other by equal and opposite
amount horizontally.  They
correspond to the two terms in eq (\ref{sum})
[shown here at $T = 0$]. 
 The resultant
$I(\chi)$, full-line,  is anomalous.  The corresponding junction energy,
being proportional to the integral of $I$ over $\chi$, 
(line with decorated with symbols) has a relative minimum at $\chi = \pi$.
%[Any $\chi$ such that $I(\chi)=0$ always 
%corresponds to an energy extremum.  $\chi$
%is a local energy minimum if $d I / d \chi > 0$.] 
This mechanism is operative so long as $T$ is not too close
to $T_c$, so that the individual terms in expression (\ref{sum})
is not strictly sinusoidal. }

%\vskip 0.3 cm
\label{fig:idea}
\end{minipage}
\end{figure}

With this, we can immediately see a mechanism for the formation
of the Josephson $\pi$-states if $\chi_{\hat p}^s \ne 0$,
see Fig \ref{fig:idea}. \cite{Yip95}
%Though Fig \ref{fig:idea} resembles the earlier experimental
%data \cite{Backhaus98}, 
It remains to 
%investigate the form of $I (\chi)$ after summing 
sum over the contributions from all
$\hat p$.  This can easily be done with the final result
\begin{eqnarray}
I_{N} &=& { \pi \over 2}  A N_f \Delta_B \int { d \Omega_{\hat p}
\over 4 \pi} | v_{fz} | \times \nonumber \\
 & & \qquad \sum_{\sigma} 
{\rm sin} \left( {\chi_{\hat p}^{\sigma} \over 2} \right ) 
{\rm tanh} \left(
{\Delta_B \over 2 T}
 {\rm cos} \left( {\chi_{\hat p}^{\sigma} \over 2} \right ) 
\right)
\end{eqnarray}
\noindent
where $\chi_{\hat p}^{\sigma} \equiv \chi - \sigma \chi_{\hat p}^s$;
$A$ the area of the pinhole, $N_f$ the density of states per spin
at the fermi energy, $v_f$ the Fermi velocity.
  To complete the calculation we only need
to find $\chi_{\hat p}^s$ for given $\hat n^{l,r}$,
with 
$\hat d^l_i(\hat p) = R_{i \mu} (\hat n^l, \theta_L) \hat p_{\mu}$
and similarly for $l \rightarrow r$. 

\begin{figure}[h]
\centerline
{ \epsfxsize=0.22\textwidth
\epsfysize = 0.37\textwidth
\rotate[r]
{ \epsfbox{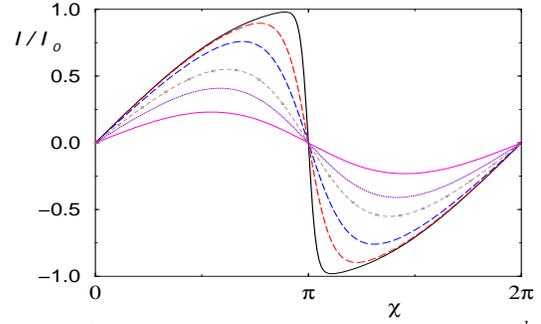} } }
\vskip 0.4 cm
\begin{minipage}{0.45\textwidth}
\caption[]{ Current-phase relationships for $\hat n^{l,r}$ both
along the normal and parallel to each other.
The temperatures are, for decreasing critical current,
 $ T/T_c =$ $ 0.1, 0.3, 0.5, 0.7, 0.8, 0.9 $.
$I_o \equiv \pi A N_f v_f \Delta_B/2$ }

%\vskip 0.3 cm
\label{fig:jpp}
\end{minipage}
\end{figure}

\begin{figure}[h]
\centerline
{ \epsfxsize=0.22\textwidth
\epsfysize = 0.37\textwidth
\rotate[r]
{ \epsfbox{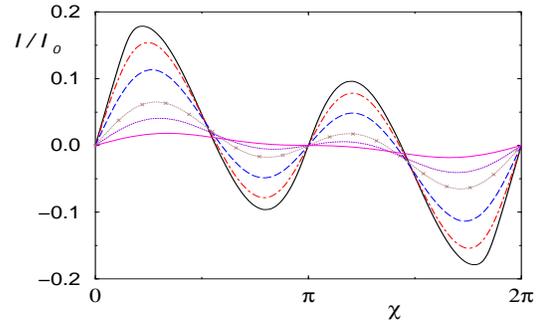} } }
\vskip 0.4 cm
\begin{minipage}{0.45\textwidth}
\caption[]{ Current-phase relationships for $\hat n^{l,r}$ both
along the normal but opposite to each other. 
The temperatures are, for decreasing critical current,
 $ T/T_c =$ $ 0.1, 0.3, 0.5, 0.7, 0.8, 0.9 $.
%$I_o \equiv \pi A N_f v_f \Delta_B/2$
 }

%\vskip 0.3 cm
\label{fig:jopp}
\end{minipage}
\end{figure}

In the absence  of any other orientation effects such
as magnetic field, $\hat n^{l,r}$ are expected to
lie along $\pm \hat z$.\cite{BC78}  If $\hat n^l = \hat n^r$ then
obviously $\hat d^l \cdot \hat d^r = 1$ hence
$\chi_{\hat p}^s = 0$  for all $\hat p$.
Our result for the current reduces to
that of an s-wave superconductor \cite{Kurki88}.
For ease of later comparison we plot the current-phase
relationship in Fig \ref{fig:jpp}. 
This is the configuration with a higher critical current.
Now consider $\hat n^l = - \hat n^r $.  Parametrizing
$\hat p$ by its azimuthal and polar angles
$(\alpha_{\hat p}, \beta_{\hat p})$, the corresponding
angles of $\hat d^{l,r}$ are obviously 
$(\alpha_{\hat p} \pm \theta_L, \beta_{\hat p})$,
One then easily gets 
$\hat d^l \cdot \hat d^r = 1 - 2 {\rm sin}^2 \beta_{\hat p}
   {\rm sin}^2 \theta_L = 1 - {15 \over 8} {\rm sin}^2 \beta_{\hat p}$.
The resultant $I(\chi)$ is as shown in Fig \ref{fig:jopp}.
This is the configuration with a lower critical current.
Except for $T$ very close to $T_c$ where $I(\chi)$ is basically
sinusoidal with a slight tilt,  $\pi$-states are evident.
These current-phase relationships resemble closely those
 obtained experimentally 
\cite{Packard99} for the ``low critical current'' state.

It is, however, known that the orientation of $\hat n$ can
be affected by a magnetic field.  The relevant terms in surface
free energies are proportional to $ - ( \hat z \cdot \hat n)^2$
and $ - (\vec H_i R_{i \mu } \hat z_{\mu} )^2 $ \cite{BC78}.  
The first term prefers $\hat n = \pm \hat z$.  However,
for sufficiently large magnetic field ( $ {> \atop \sim} 50 G$)
the second term dominates which tends to orient $\hat n$
in a direction such that the rotation $R(\hat n, \theta_L)$
rotates $\pm \hat z$ to $\hat H$.  For simplicity in the 
following we shall consider this case only. 
 Without loss of generality
we let $\hat H$ be in the $y-z$ plane and denote
its angle with the $\hat z$ axis by $\theta_H$; 
( $ 0 < \theta_H < \pi )$.  Then the possible orientations of
$\hat n$ are

\begin{eqnarray}
\left( - \sqrt{ 3 \over 5} {{\rm sin} \theta_H \over  1 + {\rm cos} \theta_H}, 
  \pm {{\rm sin} \theta_H \over  1 + {\rm cos} \theta_H} 
       \sqrt{ 1 + 4 {\rm cos} \theta_H \over 5 }, 
    \pm  \sqrt{ 1 + 4 {\rm cos} \theta_H \over 5 } \right) \nonumber
\\   
\left( + \sqrt{ 3 \over 5} {{\rm sin} \theta_H \over  1 - {\rm cos} \theta_H}, 
  \mp {{\rm sin} \theta_H \over  1 - {\rm cos} \theta_H} 
       \sqrt{ 1 - 4 {\rm cos} \theta_H \over 5 }, 
    \pm  \sqrt{ 1 - 4 {\rm cos} \theta_H \over 5 } \right) \nonumber
\end{eqnarray}

We shall use the letters A, B, C, D to denote
the different orientations of $\hat n$.  A and B exist only
for ${\rm cos} \theta_H > -{ 1 \over 4}$, whereas C and D
exist only for 
 ${\rm cos} \theta_H < { 1 \over 4}$.
At $\theta_H = 0$ the configurations A and B correspond
to $\hat n = \pm \hat z$ respectively.
A and B rotate $\hat z$ to $\hat H$ whereas
C and D rotate $-\hat z$ to $\hat H$.
For the junction
we shall denote the order parameter configurations on the two sides
by the order pairs AB etc where the letters indicate
 $\hat n^{l,r}$ respectively.  Thus if $0 < \theta_H < 0.42 \pi$
then the allowed configurations of the junction
are AA, AB, BA and BB; whereas for $ 0.42 \pi < \theta_H < 0.52 \pi$
sixteen configurations are allowed.  The current-phase
relationships of some of these configurations 
are identical by symmetry considerations alone.
A rotation of $\pi$ around the $\hat x$ axis effects the 
transformations $ A \leftrightarrow B$, $C \leftrightarrow D$
and simultaneously interchanges $l$ and $r$.  Thus {\it e.g.}
$I_{\rm AC} (\chi) = - I_{\rm DB} ( - \chi) = I_{\rm DB} (\chi)$.
In our present approximation of no surface pair breaking,
$I(\chi)$ depends only on $\hat d^{l} (\hat p) \cdot \hat d^{r} (\hat p)$.
Thus we have $I_{\rm AA} (\chi) = I_{\rm BB} (\chi)$ and 
$I_{\rm AB} (\chi) =  I_{\rm BA} (\chi)$ etc.
We are thus left with $5$ independent current-phase relationships
for ${\rm AA}$ ( $= {\rm BB} = {\rm CC} ={\rm DD})$
\cite{equalsign}
${\rm AB}$ ($={\rm BA}$), ${\rm AC}$ ($={\rm CA}={\rm BD}={\rm DB}$), 
${\rm AD}$ ( $={\rm CB}={\rm BC}={\rm DA}$) 
 and ${\rm CD}$ ($={\rm DC}$).
It turns out there is also a rather non-trivial relation between
${\rm AC}$ and ${\rm AD}$ in that for any given $\hat p$ in ${\rm AC}$,
there exists another $\hat p'$ related by rotation
about $\hat z$ such that
$\hat d^{\rm A} (\hat p) \cdot \hat d^{\rm C} (\hat p)
 = \hat d^{\rm A} (\hat p') \cdot \hat d^{\rm D} (\hat p') $
(Appendix A).
Thus $I_{\rm AC} (\chi) = I_{\rm AD} (\chi)$.
Summarizing, for given $\theta_H$ with 
$0 < \theta_H < 0.42 \pi$ there are two possible $I(\chi)$,
we shall label them AA and AB;
for $0.42 \pi < \theta_H < 0.58 \pi$ there are 
four possible $I(\chi)$'s.  We denote these
by ${\rm AA}$, ${\rm AB}$, ${\rm AC}$ and $\rm CD$.
Results for $0.58 \pi < \theta_H < \pi$ can be obtained
from those of $ 0 < \theta_H < 0.42 \pi$ by $\theta_H \rightarrow
\pi - \theta_H$.

\begin{figure}[h]
\centerline
{ \epsfxsize=0.22\textwidth
\epsfysize = 0.37\textwidth
\rotate[r]
{ \epsfbox{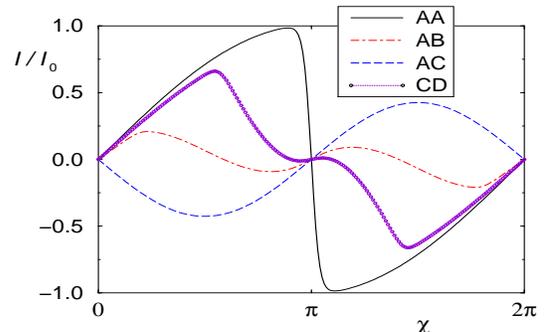} } }
\vskip 0.4 cm
\begin{minipage}{0.45\textwidth}
\caption[]{ Current-phase relationships for $\theta_H = 0.45 \pi$.
$T = 0.1 T_c$ }

%\vskip 0.3 cm
\label{fig:h0.45}
\end{minipage}
\end{figure}

As an example we show in Fig. \ref{fig:h0.45} the current-phase
relationships for these configurations at $\theta_H = 0.45 \pi$,
$T = 0.1 T_c$. $I_{\rm AA}$ is the same as that between
two s-wave superconductors since $\hat n^{l,r}$ are parallel.
AB has energy minimum at $\chi = 0$ but also a relative
minimum at $\chi = \pi$.  AC has a rather conventional shape
except for the phase shift by $\pi$, thus having
its energy minimum at $\chi = \pi$ rather than $0$.  ${\rm CD}$ has
a very weak relative minimum at $\pi$.

At $\theta_H = \pi/2$ the system possesses an extra symmetry:
a rotation of $\pi$ around the $\hat z$ axis
induces the transformations $ A \leftrightarrow {\rm C}$ and
${\rm B} \leftrightarrow {\rm D}$.  Thus at $\theta_H = \pi/2$
$I_{\rm AB}$ and $I_{\rm CD}$ merge and only three possible
$I (\chi)$ remain. (not shown)

The above provides a possible test of the hypothesis that
the $\pi$ state is the result of relative $\hat n^{l,r}$ orientations.
If one performs cool down from the normal state 
in a magnetic field (of suitable
orientation), in principle all configurations are reachable.
There should be two possible $I(\chi)$ for 
$ 0 < \theta_H < 0.42 \pi$ but at least four
for $ 0.42 \pi < \theta_H < 0.58 \pi$. 
[except $\theta_H = \pi/2$ ]

Next we consider the evolution of $I(\chi)$ as function
of $\theta_H$ for a given configuration.  
%This can be achieved experimentally by rotating the field continuously. 
 We shall in particular discuss
the case where $\theta_H$ is increased from $0$.
For AA, $\hat n^{l,r}$ remains parallel
and thus $I (\chi)$ is independent of $\theta_H$.
The result for AB is as shown in Fig \ref{fig:gab}.
Note as mentioned $\theta_H = 0$ corresponds to
$\hat n$ antiparrallel and along $\pm \hat z$.
As $\theta_H$ increases from $0$ initially
the critical current varies in a non-monotonic way 
(Appendix B); then $I(\chi)$ evolves towards
the s-wave result, reaching it at $\theta_H \approx 0.58 \pi$
where $\hat n^{l,r}$ become parallel and both along $-\hat x$.

This provides yet another test whether the $\pi$-states
observed in ref \cite{Backhaus98,Packard99} are due
to $\hat n^{l,r}$ opposite to each other.  Starting from
the configuration where the low critical current state
is  observed, if one applies
first a magnetic field along $\hat z$ of sufficient
magnitude and then rotates the magnetic field away
from the normal, provided no sudden rearrangment of
$\hat n^{l,r}$ takes place, the critical current should evolve
according to Fig \ref{fig:gab}; in particular ultimately
it should increase, the energy relative minimum at $\chi = \pi$ should become
more and more shallow and eventually disappear near $\theta_H = 0.58 \pi$. 

For completeness we also mention the $\theta_H$ 
dependences of other configurations.
$I_{\rm AC} (\chi)$ is $\theta_H$ independent
under the present approximation (Appendix A).    $I_{\rm CD} (\chi)$
can be obtained from $I_{\rm AB}$ by 
substituting $\theta_H \rightarrow \pi - \theta_H$.

Though the relative orientation between $\hat n^{l,r}$
provides a natural mechanism of $\pi$-states,
not all features observed in the Berkeley 
experiments \cite{Backhaus98,Packard99}
are consistent with the results here.
At zero magnetic field
the theory here expects $\pi$ states only for
$\hat n$ anti-parallel (and along the normal to the interface).
We therefore must identify the result of Ref \cite{Backhaus98}
as due to this configuration.
% hypothesis that the $\pi$-state
%there is due to $\hat n^{l,r}$ being oriented opposite to each other. 
%For the experimental situation of 
%Ref \cite{Backhaus98}, 
The value of $I_o$ defined in the
caption of Fig \ref{fig:jpp} corresponds to a mass current of
$\sim 2 \times 10^{-7} g/ sec$.
Thus the critical current of the "low critical current state"
at, e.g., $T = 0.28T_c$ is expected to be
only around $ 3 \times 10^{-8} g /sec $
according to Fig \ref{fig:jopp}; whereas the experimental
value is $ \sim 7 \times 10^{-8} g /sec $.\cite{Backhaus98}  
This discrepancy may be due to the finite size of the apertures
and remains to be understood.  Anyway the prediction
of strong $\hat H$ dependence of $I(\chi)$ here
can serve as an important test of the hypothesis 
that the $\pi$-states observed are due to internal
spin structure of the superfluid.

\bigskip
{\it Appendix} {\it A} -- In this Appendix we discuss
$\hat d^r (\hat p) \cdot \hat d^l (\hat p)$
for configurations AC and AD.  Obviously this
dot product is given by $\hat p_i {\cal R}_{i \mu} \hat p_{\mu}$
where ${\cal R}$ is the rotational matrix formed by
$ [  R(\hat n^r, \theta_L) ] ^{-1} R(\hat n^l, \theta_L) $.
${\cal R}$ is thus the combined action of 
$  R(\hat n^l, \theta_L) $ and then the inverse of
$  R(\hat n^r, \theta_L) $.  
First we observe that  since 
$  R(\hat n^l, \theta_L) $ rotates $ \hat z$ to $\hat H$
whereas   
$  R(\hat n^r, \theta_L) $ rotates $- \hat z$ to $\hat H$,
${\cal R}$ rotates $\hat z$ to $- \hat z$.  
From the expressions for 
$  R(\hat n^{r,l}, \theta_L) $ one can easily evaluate
the rotational angle $\Theta$ associated with ${\cal R}$ by
the formula $ {\rm Tr} {\cal R} = ( 1 + 2 {\rm cos} \Theta ) $. 
After some straightforward algebra, one can obtain 
$\Theta = \pi$.  Thus ${\cal R}$ must correspond to a
rotation of $\pi$ around an axis in the $x-y$ plane.
${\cal R}$ for AC and AD differ only by the direction
of this axis.  Thus for any given $\hat p$ for AC,
there exists another $\hat p'$ related to $\hat p$ by
a rotation around $\hat z$ such that
$\hat d^{\rm A} (\hat p) \cdot \hat d^{\rm C} (\hat p)
 = \hat d^{\rm A} (\hat p') \cdot \hat d^{\rm D} (\hat p') $
and thus their $I(\chi)$ are identical.
Also, $\theta_H$ affects only the direction of the rotational
axis for ${\cal R}$.  Thus $I(\chi)$ for these configurations
are independent of $\hat H$.

{\it Appendix} {\it B} -- Here we discuss the non-monotonic
dependence of the critical current for the configuration AB
under increasing $\theta_H$.  Using considerations along
the same lines as in Appendix A, we see that ${\cal R}$
now leaves $\hat z$ invariant and thus ${\cal R}$ must
correspond to a rotation around $\hat z$ itself.
$\Theta$ can be evaluated to be 
${\rm cos}^{-1} \left\{ {1 \over 2} \left(
{ 1 - 2 {\rm cos} \theta_H \over 1 + {\rm cos} \theta_H }\right) ^2
  -1 \right\} $.
The quantity in the bracelets and thus $\Theta$ is 
non-monotonic in $\theta_H$: at $\theta_H = 0$, 
$\Theta = {\rm cos}^{-1} (- { 7 \over 8} ) $;
at $\theta_H = \pi /3$, $\Theta$ has its maximum value
of $\pi$ but then decreases upon further increase of $\theta_H$,
reaching $\Theta = 0$ at $\theta_H = 0.58 \pi$.
Since $\Theta$ is related to the shift of the contribution of
the two different spin species
from each other, the non-monotonic behavior of $\Theta$
results in the non-monotonic dependence of the critical current
on $\theta_H$ as shown in Fig \ref{fig:gab}. 

%%%%%%%%%%%%%%%%%%%%%%%%%%%%%%%%%%%%%%%%%%%%%%%%%%%%%%%%%%%%%% 

%%%%%%%%%%%%%%%%%%%%%%%%%%%%%%%%%%%%%%%%%%%%%%%%%

%%%%%%%%%%%%%%%%%%%%%%%%%%%%%%%%%%%%%%%%%%%%
%%%%%%%%%%%%%%%%%%%%%%%%%%%%%%%%%%%%%%%%%%%%%%%%%

%%%%%%%%%%%%%%%%%%%%%%%%%%%%%%%%%%%%%%%%%%%%
%%%%%%%%%%%%%%%%%%%%%%%%%%%%%%%%%%%%%%%%%%%%%%%%%

%%%%%%%%%%%%%%%%%%%%%%%%%%%%%%%%%%%%%%%%%%%%%%%%%
\begin{figure}[h]
\centerline
{ \epsfxsize=0.22\textwidth
\epsfysize = 0.37\textwidth
\rotate[r]
{ \epsfbox{figs/gabt0.1.4ps} } }
\vskip 0.4 cm
\begin{minipage}{0.45\textwidth}
\caption[]{ Current-phase relationships for the AB configuration
as a function of $\theta_H / \pi $ given in the legend. 
$T = 0.1 T_c$.}

%\vskip 0.3 cm
\label{fig:gab}
\end{minipage}
\end{figure}

%%%%%%%%%%%%%%%%%%%%%%%%%%%%%%%%%%%%%%%%%%%%

%%%%%%%%%%%%%%%%%%%%%%%%%%%%%%%%%%%%%%%%%%%%

\end{document}